**Molecular recognition between cadherins studied by a coarse-grained model interacting with a coevolutionary potential**

Sara Terzoli and Guido Tiana*

Department of Physics and Center for Complexity and Biosystems, Università degli Studi di Milano and INFN, via Celoria 16, 20133 Milano, Italy

**Abstract**

Studying the conformations involved in the dimerization of cadherins is highly relevant to understand the development of tissue and its failure, which is associated with tumors and metastases. Experimental techniques, like X-ray crystallography, can usually report only the most stable conformations, missing minority states that could nonetheless be important for the recognition mechanism. Computer simulations could be a valid complement to the experimental approach. However, standard all-atom protein models in explicit solvent are computationally too demanding to search thoroughly the conformational space of multiple chains composed of several hundreds of amino acids. To reach this goal, we resorted to a coarse-grained model in implicit solvent. The standard problem with this kind of models is to find a realistic potential to describe their interactions. We used coevolutionary information from cadherin alignments, corrected by a statistical potential, to build an interaction potential which is agnostic of the experimental conformations of the protein. Using this model, we explored the conformational space of multi-chain systems and validated the results comparing with experimental data. We identified dimeric conformations that are sequence-specific and that can be useful to rationalize the mechanism of recognition between cadherins.

* Guido Tiana, guido.tiana@unimi.it, tel. +39-0250317221

## Introduction

Cadherins are surface proteins responsible for cell-cell recognition and adhesion[1]. They are involved in different stages of tumor progression, like in angiogenesis[2] and metastasis[3], and several germline mutations are found in solid tumors[4]. For this reason, they are important potential targets of anti-tumoral molecules.

A large number of members of the cadherin superfamily have been discovered. Particularly important for their relationship with cancer are the so-called 'classical' cadherins of type I and II, which are present only in vertebrates and which are classified according to the tissue where they were first identified. The human genome encodes 114 cadherins, like E-cadherin was found in epithelial tissues, N-cadherin in neurons, P-cadherin in placenta[5].

Classical cadherins display five extracellular (EC) domains which are structurally similar and display a significant sequence similarity, both comparing domains of the same protein type and across different types (see Fig. S1 in the Supp. Mat.). At the interface between consecutive domains it is bound a calcium ion; the EC domains in absence of calcium are more flexible[6] and the protein loses its adhesive function[7].

The adhesion between two cells is stabilized by the *trans* dimerization of the most distal EC1 domains; crystallization experiments indicate that *trans* dimerization occurs through the swap of their N-termini[8]. Crystal structures of EC1-EC2 domains mutated at the N-termini to prevent domain swapping show that another, X-shaped, *trans* dimeric conformation is possible. Destabilization of the X-dimer by mutating a residue are the junction between EC1 and EC2 slows down the domain-swapping event, qualifying the X-dimer as an on-pathway intermediate[9].

In many *in vivo* conditions, classical cadherins are homophilic, in the sense that cells expressing the same cadherin associate, while those expressing different cadherins segregate[10]. This property is at the basis of cellular binding specificity in tissues and is critical in the correct development of organisms, as ectopic expression of cadherins leads to morphological defects[11]. However, the homophilic effect *in vivo* does not seem to be a straightforward consequence of the affinity between cadherins of the same type. In fact, analytical-ultracentrifugation and surface-plasmon-resonance experiments of purified EC1-EC2 domains do not provide dissociation constants that reflect the homophilic relations observed *in vivo*[12,13]. Although some physical models have been proposed to explain homophilic interactions in systems composed of two cellular types, expressing different cadherins[14,15], they cannot explain all aspects of cellular sorting[16,17] and cannot be easily extended to the case of many cell types. Thus, the molecular binding code remains poorly understood and indeed it requires further investigation[18].

Computational methods could in principle complement the available experimental data to give an atomic-level description, analogously to what crystal structures do, but also describing the conformational changes and the fluctuations among multiple states associated with the molecular recognition between cadherins. The main problem in this respect is that the system one wishes to simulate, for example that composed of two pairs of EC1-EC2 domains, has a molecular weight of ~50 kDa and thus it is huge from the point of view of standard atomistic simulations in explicit solvent.

Coarse-grained models based on experimental data can be useful in this context. By describing the protein system in implicit solvent and giving a united-atom representation of some atomic groups, they allow computers to sample reasonably fast the conformational space of the system. Defining an interaction potential based on experimental data, within the framework of the principle of maximum entropy, guarantees the realism of the model and minimizes the risk of introducing subjective bias in the description of the system[19].

In the present work, we employed a coevolutionary interaction potential[20], calculated from the set of homologous sequences of the cadherin superfamily. In brief, a coevolutionary potential describes the interaction between amino acids in a protein in such a way to predict the correct correlations between mutations in the alignment of homologs, as obtained from the PFAM database[21]. This kind of modelling has proven efficient in predicting the native conformation protein monomers[20] and dimers[22], of their conformational fluctuations[23,24], and of the effect of mutations in protein stability[25,26].

We first showed that the model is able to reproduce several experimental data observed for cadherins of different type. Then, we sampled the conformational space of pairs of cadherins of the same kind and of different kind to identify the sequence-dependent dimeric conformations that could be relevant for the mechanism of molecular recognition.

**Methods**

Protein chains were modelled with a united-atom representation in implicit solvent, similar to others commonly employed in the literature[27,28]. Each amino acid is described[29] by the positions of its N, CA and C atom, and by that of another bead which represents the whole side chain, and which is set in the position of its center of mass (see Fig. S2 in the Supp. Mat.). Bond lengths and angles are maintained fixed at the values defined by the initial conformation.

The interaction energy of the system is defined as

$$U = \sum_{i<j} J_{HC}\, \theta\left(R_{HC} - |r_i - r_j|\right) + \sum_{is<js} J_{is,js}\, \theta\left(|r_{is} - r_{js}| - R\right) + U_{dih}, \qquad (1)$$

where *i* and *j* run on all atoms, the function $\theta(x)$ is a step contact function which takes the values 1 if $x \geq 0$ and 0 if $x < 0$, $R_{HC}$ is the hard-core radius, the hard-core energy $J_{HC} \to +\infty$, *is* and *js* run on the side-chain atoms, $J_{is,js}$ is the interaction matrix and *R* is the interaction range. We set $R = 8.5$Å and $R_{HC} = 2$Å for all atoms.

The interaction matrix was obtained from a coevolutionary model corrected by a statistical potential. The coevolutionary model takes the alignment of homologs and returns a tensor $J_{I,K}(\sigma,\tau)$ of interaction energies between any residue of type $\sigma$ at position *I* and any residue of type $\tau$ at position *K*. This is calculated within the pseudolikelihood approximation[30] using the code described in ref. 31 (see also Sect. S1 in the Supp. Mat.). In the standard procedure[32], the pseudolikelihood is maximized under the constraint of a $l_2$ regularizer of kind $\alpha \sum_{ij} \left( J_{I,K}(\sigma,\tau) \right)^2$ meant to correct finite-size effects. An important ingredient of the present model is the use of a different type of $l_2$ regularizer, namely

$$\alpha \sum_{ij} \left( J_{I,K}(\sigma,\tau) - \varepsilon(\sigma,\tau) \right)^2, \qquad (2)$$

where $\epsilon(\sigma,\tau)$ is a statistical potential[33] obtained from the frequency of contacts in known protein structures. This is a system-independent potential calculated as

$$\varepsilon(\sigma,\tau) = -\log f(\sigma,\tau) + \varepsilon_0, \qquad (3)$$

where $f(\sigma,\tau)$ are the frequency of contacts between the side chains of the non-redundant set of proteins of the pdb (with non-redundancy threshold of $10^{-7}$). The goal of this regularizer is to provide a *a priori* knowledge of the interactions in the system, especially useful in the case of amino acids that appear with poor statistics or that coevolve under the effect of biological constraints other than the stability of the protein, effect that would produce falsely strongly interacting pairs[34]. The regularizer also avoids the problem of choosing a guage for the maximum pseudolikelihood problem.

After calculating the tensor $J_{i,j}(\sigma,\tau)$ of interacting energies between any possible pairs residue $\sigma$ and $\tau$ that can appear at sites *i* and *j* of the protein, we projected it onto the sequence of interest $\{\sigma_i\}$, obtaining the matrix $J_{i,j}(\sigma_i, \sigma_j)$. This is then normalized by $J_0 = \left[\sum_{i<j} J_{ij}(\sigma_i, \sigma_j)^2 - \left(\sum_{i<j} J_{ij}(\sigma_i, \sigma_j)\right)^2\right]^{1/2}$ in order to set the scale of two-body interactions to 1.

The *trans* interaction between identical proteins is set using the same parameters of the corresponding *cis* interaction.

The last term in the potential of Eq. (1) depends on the dihedrals of the backbone of the protein, in the form

$$U_{dih} = -\varepsilon_{dih} \sum_i \left( w_i^{\alpha} exp\left[ -\frac{(\phi_i - \phi_{i\alpha}^0)^2}{2\sigma_{i\alpha}^2} \right] + w_i^{\beta} exp\left[ -\frac{(\phi_i - \phi_{i\beta}^0)^2}{2\sigma_{i\beta}^2} \right] \right), \qquad (4)$$

where $\phi_i$ are the Ramachandran dihedrals (i.e., alternatively $\varphi$ and $\psi$), $w_i^{\alpha}$ and $w_i^{\beta}$ are the sequence-dependent propensities of being in $\alpha$ and in $\beta$ structure, respectively, as predicted by PsiPred[35], $\phi_{i\alpha}^0$ and $\phi_{i\beta}^0$ are the typical dihedrals associated with $\alpha$ (-63° for $\varphi$ and -44° for $\psi$) and $\beta$ (-105° for $\varphi$ and -140° for $\psi$) structures, and we set $\sigma_{i\alpha} = 30^o$ and $\sigma_{i\beta} = 40^o$.

Thus, the potential depends on three energy (meta)parameters, namely $\alpha$, $\varepsilon_0$ and $\varepsilon_{dih}$. We explored the space of parameters in the case of small proteins (see Sect. S2 in the Supp. Mat.) and found the optimal values $\alpha=10^{-5}$, $\varepsilon_0=-1$ and $\varepsilon_{dih}=90$.

Simulations were performed with a Metropolis Monte Carlo (MC) algorithm, using as elementary moves multiple flips, pivots and roto-translations of the center of mass of connected systems of chains[36]. Parallel-tempering simulations[37] were performed trying an exchange between replicas of adjacent temperatures every 1000 MC steps.

Since protein domains tend to attract each other quite strongly, we performed an umbrella sampling[38] to equilibrate the system more efficiently. The interaction between different chains is rescaled by a factor $k<1$. The correct equilibrium probabilities $p(r)$ of the conformations of the system are then recovered *a posteriori* from the simulated probabilities $p_k(r)$ as

$$p(r) = p_k(r) \frac{exp[\beta(k-1)E_{trans}(r)]}{\langle exp[\beta(k-1)E_{trans}]\rangle_k}, \qquad (5)$$

where $E_{trans}(r)$ is the rescaled interaction between the chains, $\beta$=1/T is the inverse temperature (we set Boltzmann's constant to 1, expressing temperatures in energy units) and the angular brackets with the subscript *k* at the denominator indicate the average obtained by the simulation. In this way, we could make the simulation faster, having a larger $k_{off}$ between the two chains, and recover the correct equilibrium properties *a posteriori*.

## Results

*Coevolutionary potentials reproduce the native state of monomeric EC1*

As a first step, we simulated the dynamics of EC1 of N-, E-, and P-cadherin (residues 1-99) in conditions of infinite dilution, where the protein is monomeric. We used as putative monomeric reference conformations the crystallographic structures 2qvi (for N-cadherin), 2o72 (for E-cadherin) and 4zmz (for P-cadherin). At low temperature, the average RMSD calculated from parallel-tempering simulations with respect to the crystallographic conformations is ≈ 0.5 nm for N- and E-cadherin and ≈ 0.7 nm for P-cadherin (see Fig. 1a). This is comparable with that of the proteins previously studied (cf. Sect. S2 in the Supp. Mat.). The calculated contact-probability maps are also native-like (cf. Fig. 1f). These two facts suggest that the minimum requirement for the model to be useful, that is to have the experimental conformations as low-temperature equilibrium states, is met. It is important to stress that, unlike structure-based models[39], here the model is agnostic of the native conformation of the proteins. Moreover, the use of the regularizer given by Eq. (2) based on a system-independent statistical potential seems important to obtain a realistic potential, since not using it increases markedly the RMSD of the protein (cf. Fig. S3).

At varying temperature, all three proteins display a main transition at temperature T≈5 (in energy units, see Fig. 1c) between a native and a denatured state. The transition, as described by the model, appears as poorly cooperative; although we are not aware of calorimetric studies of the EC1 domains of cadherins, it is likely that, as in most implicit-solvent models[40], this is an artifact associated with the use of reduced degrees of freedom.

The thermal fluctuations of the residues display similar patterns in the three monomeric cadherins (see Fig. 1d), but their relative widths are protein-dependent (see Fig. 1e): E-cadherin displays larger fluctuations in the proximal region (i.e., that linked to EC2 in the full complex), while P-cadherin fluctuates more in the distal region, and N-cadherin behaves in an intermediate way (cf. Fig. 1e). The fluctuations of the residues of N-cadherin display a significant correlation (Pearson's r=0.47, p-value$<10^{-5}$) with the b-factors of its crystallographic structure (see dashed line in Fig. 1d; note that 1NCJ is the only available structure of a monomeric EC1 domain of a classical cadherin).

*The model correctly predicts the dimeric structures of EC1 of N-cadherin*

The next step was to simulate two EC1 domains, that is the system studied in the original work of Shapiro and coworkers[8]. The two chains are put in a spherical box of volume $V \approx 3.2 \cdot 10^4$ nm$^3$, corresponding to a concentration of $\approx$ 100 μM. The fraction $f_B$ of conformations displaying inter-chain contacts is displayed in Fig. 2f. The experimental $k_D$ obtained at room temperature from analytical size-exclusion chromatography on EC1 alone is 166 μM[41], corresponding to a $f_B=0.3$ in the simulation volume. This allowed us to set the simulation temperature T=3.6 (in energy units) as that corresponding to room temperature (see red arrow in Fig. 2f).

The average contact map simulated at T=3.6 is displayed in Fig. 2a, together with the contact maps of the two alternative crystallographic structures found for EC1[8] describing the domain-swapped "strand dimer" (pdb code 1NCI) and the "adhesion dimer" (pdb code 1NCH). The binding of the two monomers does not perturb their internal structure, the intra-chain contacts remaining the same as the crystallographic ones, and the average RMSD remaining ≈0.5 as in the monomeric case (cf. Fig. 1a).

The simulations also display several inter-chain contacts of varying stability. Two sets of contacts, marked with green and red boxes in Fig. 2a, correspond to the contacts of the strand dimer and of the adhesion dimer, respectively. Another set of contacts, formed with non-negligible probability, cannot be explained by the available crystallographic structures.

The simulated conformations were clustered based on their mutual similarity and the most representative conformations are shown in Fig. 2b-e, together with the associated probabilities. In approximately one third of the conformations the two chains are disjoint (Fig. 2b); in another third, they display a conformation similar to that of the adhesion dimer (fig. 2c), making the inter-chain contacts marked with red boxes in Fig. 2a. In 13% of the sampled conformations, the tryptophans 2W of each chain is in contact with the hydrophobic pocket of the other chain (see Fig. 2e and the green-boxed contacts in Fig. 2a). A 11% of the conformations populate a dimeric conformation (Fig. 2d) which is not similar to any available crystallographic structure, while the remaining 17% of the conformations populate dimeric structures that cannot be easily clustered into well-defined groups.

*Simulated EC1-2 fluctuate among different conformations, including X- and domain-swapped dimers*

Simulations of two copies of the chain composed of the EC1-2 domain are carried out for $10^9$ MC steps for N-, E- and P-cadherin, rescaling the inter-chain interactions as described in the Methods section. All proteins display a very heterogeneous set of dimeric conformations. A clustering analysis, whose results are reported in Fig. 3, reports that the three proteins can assemble in many possible conformations and among them there are conformations resembling the swap dimer and the X-dimer, although with a probability lower than expected. In several conformations the two chains are side-by-side or display interactions between their proximal ends.

In the case of E-cadherin[42], the average contact map is displayed in Fig. 4a. Of the 45 clusters of contacts with probability larger than 0.1 which are apparent in the contact map, 7 are those that stabilize the strand dimer (marked with green boxes, cf. also Fig. 4b) and 15 are those that stabilize the X-dimer (purple boxes, cf. also Fig. 4c). The remaining contacts cannot be explained from the crystallographic structures, but result from other conformations displayed in Fig. 3. However, 18 of these unexplained contacts involve all residues that are known to be associated with mutations observed in tumoral cells that decrease cell-cell adhesion and induce metastasis[4].

The contact maps simulated for P- and N-cadherins display overall less clusters of contacts (cf. Fig. S6 in the Supp. Mat). N-cadherin has 26 clusters, 6 of them associated with the strand dimer, 11 with the X-dimer and 10 with other conformations displayed in Fig. 3. P-cadherin[43] has 21 clusters, 5 of them associated with the strand dimer, 11 with the X-dimer and 6 with other conformations.

For E-cadherin one can also compare the results of simulations with those of double electron-electron resonance (DEER), which is able to measure the distribution of distances between labelled side chains within 6 nm. E-cadherin labelled at residues 73-75 and 114-116 display a double peak between 4.0 and 4.5 nm, interpreted as arising from the strand dimer[13]. Our simulations display a similar double peak (cf. Fig. 4d) as the result of the contribution of all conformations displayed in Fig. 3.

On the other hand, the simulated distances between residues 135 of the two chains match poorly the results obtained by DEER (cf. Fig. S7 in the Supp. Mat.), most likely because the angle between the axes of the two chains is strongly affected by the coarse graining of the model.

*Simulations of heterophilic complexes*

Further simulations were carried out with pairs of different EC1-2 domains, to simulate heterophilic interactions. The average contact maps of the hybrid systems are displayed in Fig. 5a-c. The system

E-N populates in a detectable way the swapped and the X dimer, the system P-N only the X dimer, while E-P none of the two. However, all systems populate multiple dimeric conformations, most of which are system dependent.

The residues participating in the dimeric contacts are highlighted in the alignment displayed in Fig. 5d. It is apparent that residue participating to the swapped dimer are more conserved than those participating to the X dimer, and those participating to the other types of dimers are even less conserved (cf. also Fig. 5e). As a consequence, the contacts which stabilize these other types of dimers are more specific than those in swapped and X dimers.

The dissociation constants obtained from the simulations for the homophilic and heterophilic cases respect in most cases the order of those obtained experimentally from analytical ultracentrifugation and plasmon resonance analysis[13], see Fig. 5f. The major difference is in the fact that the N-E complex is very weak ($k_D$>100 $\mu$M) in the simulations while should be of the same order of magnitude as the P-E complex (~50 $\mu$M).

**Discussion**

The atomic-scale picture we have of the encounter mechanism between cadherins is essentially based on the crystal structures of the wild-type and of mutant proteins[8,9,13,42]. The behavior of cadherins in solution, and even more *in vivo*, could be more complex than the static and homogeneous situation observed in crystals. Unfortunately, for such a large system as an assembly of cadherins, there are few experimental techniques that can report indirect conformational data in solution [13,44–47], leaving behind the problem of turning these data into a structural understanding of their recognition mechanism.

Simulating the mutual search and binding of multiple cadherins with computational techniques can be a way to obtain details that can complement experimental data and describe all the conformations involved in the mechanism of molecular recognition at atomic level. The main problems in pursuing this approach with standard molecular-dynamics simulations is that one has to deal with a system that on the scale of computer calculations is large (~50 kDa for the EC1-2 dimer) and takes a long time to bind (~1 s for E-cadherin[44]).

Coarse-grained models, combined with advanced sampling algorithms, can be useful to study this kind of systems. Using a united-atom representation, in which each amino acid is represented by 4 atoms and the solvent is treated implicitly, we could sample at equilibrium the conformational space of two copies of the EC1-2 domains with advanced Monte Carlo algorithms in few days of computational time.

The main problem with computational models of biomolecules, and with coarse-grained model in particular, is to build a realistic interaction potential. A strategy that is gaining popularity is to build potentials based on available experimental data in the framework of the principle of maximum entropy, and then to validate the model with independent data[48,49].

A particularly abundant set of data available for proteins, and for cadherins in particular, is sequence data, in the form of alignments of homologous protein sequences. Coevolutionary analysis is a way to extract a contact potential from these data, finding the most likely potential that could have produced the available alignment as the result of natural evolution. There are several implementations of this idea[24,30], all of them giving comparable results[34]. Coevolutionary potentials proved useful to predict the native state of single-domain proteins[24,30], of their energy profile[23], of protein-protein interactions[50] and of the thermodynamic effect of point mutations[25,26].

In the present work, we applied a coevolutionary potential to the problem of molecular recognition between cadherins. The coevolutionary potential was corrected with a system-independent statistical potential, obtained from the contact probabilities obtained from the whole pdb. This appears to be an important step, because it corrects those terms of the coevolutionary potential associated with poor statistics in the cadherin alignments, and then otherwise affected by large noise.

We validated the model in several ways, also estimating what are its limitations. First, we verified that the monomeric system displays at low temperature a unique native conformation compatible with the crystallographic one. This was tested for the EC1 domain of three different cadherins and for two other small proteins used as independent control. The positive result we obtained is not straightforward because, unlike structure-based models widely used to study conformational changes in proteins[51], we never used any information about the native conformation of the system during the construction of the potential. The accuracy with which we could simulate the native state of monomers is somewhat worse than the experimental resolution of X-ray structures, being quantified by an RMSD of the order of 0.5-0.7 nm. This is due to the united-atom modelling of amino acids, that is required to fasten the simulation but that does not lead to a perfect packing of side chains.

Moreover, we compared the simulated trajectories with the experimental b-factor, with the results of analytical size-exclusion chromatography, with double electron-electron resonance experiments and with the dissociation constants obtained by analytical ultracentrifugation and plasmon resonance analysis. Also the dimeric structures generated by the simulations were compared to the crystallographic structures available for the N-, E- and P-cadherin. Interestingly, in all the three simulations we find structures similar to the swap-dimer and the X-dimer that was identified in crystals. Since wild-type cadherins crystallize into swap-dimers, one would have expected that this structure displayed the largest population fraction in the simulation. One reason why this is not the case could be that the swap dimer is easier to crystallize, and thus the experiments selects only one of the possible conformers. Of course, the coarse-graining of the model also affects the entropy of the system, on which the probability depends exponentially.

One can thus wonder whether the heterogeneous set of binding modes of the EC1-2 dimers observed in the simulations is realistic or just an artifact of the model. An element pointing towards their realism is that all pathogenic mutations observed in E-cadherin[4] affect the interface of these conformations, leading *in vivo* to a diminished adhesion and increased migration propensity of the cells. This fact suggests that the dimeric conformers we found may play a role in the overall binding mechanism.

This fact becomes more evident when simulating heterophilic interactions in hybrid systems composed of different types of cadherins. Also in this case we could observe conformations resembling the swapped and the X dimer, in addition to a complex set of other dimeric conformations. Interestingly, while the residues that stabilize swapped and X dimers are quite independent on the kind of cadherin, and thus weakly specific, the other dimers interact through contacts that are much more system dependent. One can then speculate that the multiplicity of dimeric conformations different than the swapped and the X dimer play a role in the selective molecular recognition between cadherins.

**Conclusions**

We proposed a coarse-grained model interacting through a potential based on the coevolutionary analysis of homologous proteins to study the molecular recognition between molecules that are too large to be studied with standard molecular-dynamics simulations. The model relies on sequence data and is agnostic of the structural properties of the protein. It was validated comparing the results of Monte Carlo simulations with experimental data of various type, giving good agreement except for the relative populations of the different types of dimeric structures that depend exponentially on the energies that define the model, and are thus quite sensitive to them.

A thorough sampling of the conformational space of dimers composed of pairs of EC1-2 domains of E-, P- and N-cadherins show that besides the known swapped and X dimers, the systems populate multiple other dimeric conformations, that are more sequence-dependent and thus could play an important role in the selectivity of molecular recognition between cadherins.

## References

(1) Takeichi, M. Cadherin Cell Adhesion Receptors as a Morphogenetic Regulator. *Science* **1991**, *251* (5000), 1451–1455.

(2) Blaschuk, O. W. N-Cadherin Antagonists as Oncology Therapeutics. *Philos. Trans. R. Soc. Lond. B. Biol. Sci.* **2015**, *370* (1661), 20140039.

(3) Berx, G.; Becker, K. F.; Höfler, H.; van Roy, F. Mutations of the Human E-Cadherin (CDH1) Gene. *Hum. Mutat.* **1998**, *12* (4), 226–237.

(4) Petrova, Y. I.; Schecterson, L.; Gumbiner, B. M. Roles for E-Cadherin Cell Surface Regulation in Cancer. *Mol. Biol. Cell* **2016**, *27* (21), 3233–3244.

(5) Tiwari, P.; Mrigwani, A.; Kaur, H.; Kaila, P.; Kumar, R.; Guptasarma, P. Structural-Mechanical and Biochemical Functions of Classical Cadherins at Cellular Junctions: A Review and Some Hypotheses. *Adv. Exp. Med. Biol.* **2018**, *1112* (5865), 107–138.

(6) Häussinger, D.; Ahrens, T.; Sass, H.-J. J.; Pertz, O.; Engel, J.; Grzesiek, S. Calcium-Dependent Homoassociation of E-Cadherin by NMR Spectroscopy: Changes in Mobility, Conformation and Mapping of Contact Regions. *J. Mol. Biol.* **2002**, *324* (4), 823–839.

(7) Sedar, A. W.; Forte, J. G. Effects of Calcium Depletion on the Junctional Complex between Oxyntic Cells of Gastric Glands. *J. Cell Biol.* **1964**, *22* (1), 173–188.

(8) Shapiro, L.; Fannon, A. M.; Kwong, P. D.; Thompson, A.; Lehmann, M. S.; Grübel, G.; Legrand, J. F.; Als-Nielsen, J.; Colman, D. R.; Hendrickson, W. A. Structural Basis of Cell-Cell Adhesion by Cadherins. *Nature* **1995**, *374* (6520), 327–337.

(9) Harrison, O. J.; Bahna, F.; Katsamba, P. S.; Jin, X.; Brasch, J.; Vendome, J.; Ahlsen, G.; Carroll, K. J.; Price, S. R.; Honig, B.; et al. Two-Step Adhesive Binding by Classical Cadherins. *Nat. Struct. Mol. Biol.* **2010**, *17* (3), 348–357.

(10) Duguay, D.; Foty, R. A.; Steinberg, M. S. Cadherin-Mediated Cell Adhesion and Tissue Segregation: Qualitative and Quantitative Determinants. *Dev. Biol.* **2003**, *253* (2), 309–323.

(11) Detrick, R. J.; Dickey, D.; Kintner, C. R. The Effects of N-Cadherin Misexpression on Morphogenesis in Xenopus Embryos. *Neuron* **1990**, *4* (4), 493–506.

(12) Katsamba, P.; Carroll, K.; Ahlsen, G.; Bahna, F.; Vendome, J.; Posy, S.; Rajebhosale, M.; Price, S.; Jessell, T. M.; Ben-Shaul, A.; et al. Linking Molecular Affinity and Cellular Specificity in Cadherin-Mediated Adhesion. *Proc. Natl. Acad. Sci. U. S. A.* **2009**, *106* (28), 11594–11599.

(13) Vendome, J.; Felsovalyi, K.; Song, H.; Yang, Z.; Jin, X.; Brasch, J.; Harrison, O. J.; Ahlsén, G.; Bahna, F.; Kaczynska, A.; et al. Structural and Energetic Determinants of Adhesive Binding Specificity in Type I Cadherins. *Proc. Natl. Acad. Sci. U. S. A.* **2014**, *111* (40), E4175–E4184.

(14) Steinberg, M. S. Reconstruction of Tissues by Dissociated Cells. *Science* **1963**, *141* (3579), 401–408.

(15) Harris, A. K. Is Cell Sorting Caused by Differences in the Work of Intercellular Adhesion? A Critique of the Steinberg Hypothesis. *J. Theor. Biol.* **1976**, *61* (2), 267–285.

(16) Pawlizak, S.; Fritsch, A. W.; Grosser, S.; Ahrens, D.; Thalheim, T.; Riedel, S.; Kießling, T. R.; Oswald, L.; Zink, M.; Manning, M. L.; et al. Testing the Differential Adhesion Hypothesis across the Epithelial–mesenchymal Transition. *New J. Phys.* **2015**, *17* (8), 083049.

(17) Krieg, M.; Arboleda-Estudillo, Y.; Puech, P.-H.; Käfer, J.; Graner, F.; Müller, D. J.; Heisenberg, C.-P. Tensile Forces Govern Germ-Layer Organization in Zebrafish. *Nat. Cell Biol.* **2008**, *10* (4), 429–436.

(18) Green, J. B. A. Sophistications of Cell Sorting. *Nat. Cell Biol.* **2008**, *10* (4), 375–377.

(19) Jaynes, E. T. Information Theory and Statistical Mechanics. *Phys. Rev.* **1957**, *106*, 620–630.

(20) Morcos, F.; Pagnani, A.; Lunt, B.; Bertolino, A.; Marks, D. S.; Sander, C.; Zecchina, R.; Onuchic, J. N.; Hwa, T.; Weigt, M. Direct-Coupling Analysis of Residue Coevolution Captures Native Contacts across Many Protein Families. *Proc. Natl. Acad. Sci. U. S. A.* **2011**, *108* (49), E1293-301.

(21) Finn, R. D.; Coggill, P.; Eberhardt, R. Y.; Eddy, S. R.; Mistry, J.; Mitchell, A. L.; Potter, S. C.; Punta, M.; Qureshi, M.; Sangrador-Vegas, A.; et al. The Pfam Protein Families Database: Towards a More Sustainable Future. *Nucleic Acids Res.* **2016**, *44* (D1), D279-85.

(22) dos Santos, R. N.; Morcos, F.; Jana, B.; Andricopulo, A. D.; Onuchic, J. N. Dimeric Interactions and Complex Formation Using Direct Coevolutionary Couplings. *Sci. Rep.* **2015**, *5* (1), 13652.

(23) Sutto, L.; Marsili, S.; Valencia, A.; Gervasio, F. L. From Residue Coevolution to Protein Conformational Ensembles and Functional Dynamics. *Proc. Natl. Acad. Sci. U. S. A.* **2015**, *112* (44), 13567–13572.

(24) Morcos, F.; Jana, B.; Hwa, T.; Onuchic, J. N. Coevolutionary Signals across Protein Lineages Help Capture Multiple Protein Conformations. *Proc. Natl. Acad. Sci. U. S. A.* **2013**, *110* (51), 20533–20538.

(25) Lui, S.; Tiana, G. The Network of Stabilizing Contacts in Proteins Studied by Coevolutionary Data. *J. Chem. Phys.* **2013**, *139* (15), 155103.

(26) Contini, A.; Tiana, G. A Many-Body Term Improves the Accuracy of Effective Potentials

Based on Protein Coevolutionary Data. *J. Chem. Phys.* **2015**, *143* (2), 25103.

(27) Liwo, A.; Khalili, M.; Scheraga, H. A. Ab Initio Simulations of Protein-Folding Pathways by Molecular Dynamics with the United-Residue Model of Polypeptide Chains. *Proc. Natl. Acad. Sci. U. S. A.* **2005**, *102* (7), 2362–2367.

(28) Kmiecik, S.; Kolinski, A. Characterization of Protein-Folding Pathways by Reduced-Space Modeling. *Proc. Natl. Acad. Sci.* **2007**, *104* (30), 12330–12335.

(29) Voegler Smith, A.; Hall, C. K. Alpha-Helix Formation: Discontinuous Molecular Dynamics on an Intermediate-Resolution Protein Model. *Proteins Struct. Funct. Genet.* **2001**, *44* (3), 344–360.

(30) Ekeberg, M.; Lövkvist, C.; Lan, Y.; Weigt, M.; Aurell, E. Improved Contact Prediction in Proteins: Using Pseudolikelihoods to Infer Potts Models. *Phys. Rev. E, Stat. nonlinear, soft matter Phys.* **2013**, *87* (1), 620630.

(31) Fantini, M.; Malinverni, D.; De Los Rios, P.; Pastore, A. New Techniques for Ancient Proteins: Direct Coupling Analysis Applied on Proteins Involved in Iron Sulfur Cluster Biogenesis. *Front. Mol. Biosci.* **2017**, *4*, 40.

(32) Ekeberg, M.; Hartonen, T.; Aurell, E. Fast Pseudolikelihood Maximization for Direct-Coupling Analysis of Protein Structure from Many Homologous Amino-Acid Sequences. *J. Comput. Phys.* **2014**, *276*, 341–356.

(33) Miyazawa, S.; Jernigan, R. Estimation of Effective Interresidue Contact Energies from Protein Crystal Structures: Quasi-Chemical Approximation. *Macromolecules* **1985**, *18*, 534–552.

(34) Franco, G.; Cagiada, M.; Bussi, G.; Tiana, G. Statistical Mechanical Properties of Sequence Space Determine the Efficiency of the Various Algorithms to Predict Interaction Energies and Native Contacts from Protein Coevolution. *Phys. Biol.* **2019**, *16* (4), 046007.

(35) McGuffin, L. J.; Bryson, K.; Jones, D. T. The PSIPRED Protein Structure Prediction Server. *Bioinformatics* **2000**, *16* (4), 404–405.

(36) Tiana, G.; Villa, F.; Zhan, Y.; Capelli, R.; Paissoni, C.; Sormanni, P.; Heard, E.; Giorgetti, L.; Meloni, R. MonteGrappa: An Iterative Monte Carlo Program to Optimize Biomolecular Potentials in Simplified Models. *Comput. Phys. Commun.* **2014**, *186*, 93–104.

(37) Swendsen, R. H.; Wang, J. S. Replica Monte Carlo Simulation of Spin Glasses. *Phys. Rev. Lett.* **1986**, *57* (21), 2607–2609.

(38) Torrie, G. M.; Valleau, J. P. Nonphysical Sampling Distributions in Monte Carlo Free-Energy Estimation: Umbrella Sampling. *J. Comput. Phys.* **1977**, *23* (2), 187–199.

(39) Oliveira, L. C.; Schug, A.; Onuchic, J. N. Geometrical Features of the Protein Folding Mechanism Are a Robust Property of the Energy Landscape: A Detailed Investigation of Several Reduced Models. *J. Phys. Chem. B* **2008**.

(40) Chan, H. S. Modeling Protein Density of States: Additive Hydrophobic Effects Are Insufficient for Calorimetric Two-State Cooperativity. *Proteins Struct. Funct. Genet.* **2000**, *40* (4), 543–571.

(41) Davila, S.; Liu, P.; Smith, A.; Marshall, A. G.; Pedigo, S. Spontaneous Calcium-Independent Dimerization of the Isolated First Domain of Neural Cadherin. *Biochemistry* **2018**, *57* (45), 6404–6415.

(42) Parisini, E.; Higgins, J. M. G.; Liu, J.; Brenner, M. B.; Wang, J. The Crystal Structure of Human E-Cadherin Domains 1 and 2, and Comparison with Other Cadherins in the Context of Adhesion Mechanism. *J. Mol. Biol.* **2007**, *373* (2), 401–411.

(43) Kudo, S.; Caaveiro, J. M. M.; Tsumoto, K. Adhesive Dimerization of Human P-Cadherin Catalyzed by a Chaperone-like Mechanism. *Structure* **2016**.

(44) Li, Y.; Altorelli, N. L.; Bahna, F.; Honig, B.; Shapiro, L.; Palmer, A. G. Mechanism of E-Cadherin

Dimerization Probed by NMR Relaxation Dispersion. *Proc. Natl. Acad. Sci. U. S. A.* **2013**, *110* (41), 16462–16467.

(45) Kudo, S.; Caaveiro, J. M. M.; Goda, S.; Nagatoishi, S.; Ishii, K.; Matsuura, T.; Sudou, Y.; Kodama, T.; Hamakubo, T.; Tsumoto, K. Identification and Characterization of the X-Dimer of Human P-Cadherin: Implications for Homophilic Cell Adhesion. *Biochemistry* **2014**, *53* (11), 1742–1752.

(46) Tariq, H.; Bella, J.; Jowitt, T. A.; Holmes, D. F.; Rouhi, M.; Nie, Z.; Baldock, C.; Garrod, D.; Tabernero, L. Cadherin Flexibility Provides a Key Difference between Desmosomes and Adherens Junctions. *Proc. Natl. Acad. Sci. U. S. A.* **2015**, *112* (17), 5395–5400.

(47) Bajpai, S.; Correia, J.; Feng, Y.; Figueiredo, J.; Sun, S. X.; Longmore, G. D.; Suriano, G.; Wirtz, D. Alpha-Catenin Mediates Initial E-Cadherin-Dependent Cell-Cell Recognition and Subsequent Bond Strengthening. *Proc. Natl. Acad. Sci. U. S. A.* **2008**, *105* (47), 18331–18336.

(48) Pitera, J. W.; Chodera, J. D. On the Use of Experimental Observations to Bias Simulated Ensembles. *J. Chem. Theory Comput.* **2012**, *8* (10), 3445–3451.

(49) Tiana, G.; Giorgetti, L. Integrating Experiment, Theory and Simulation to Determine the Structure and Dynamics of Mammalian Chromosomes. *Curr. Opin. Struct. Biol.* **2018**, *49*, 11–17.

(50) Gueudré, T.; Baldassi, C.; Zamparo, M.; Weigt, M.; Pagnani, A. Simultaneous Identification of Specifically Interacting Paralogs and Interprotein Contacts by Direct Coupling Analysis. *Proc. Natl. Acad. Sci. U. S. A.* **2016**, *113* (43), 12186–12191.

(51) Noel, J. K.; Whitford, P. C.; Sanbonmatsu, K. Y.; Onuchic, J. N. SMOG@ctbp: Simplified Deployment of Structure-Based Models in GROMACS. *Nucleic Acids Res.* **2010**, *38*, W657-61.

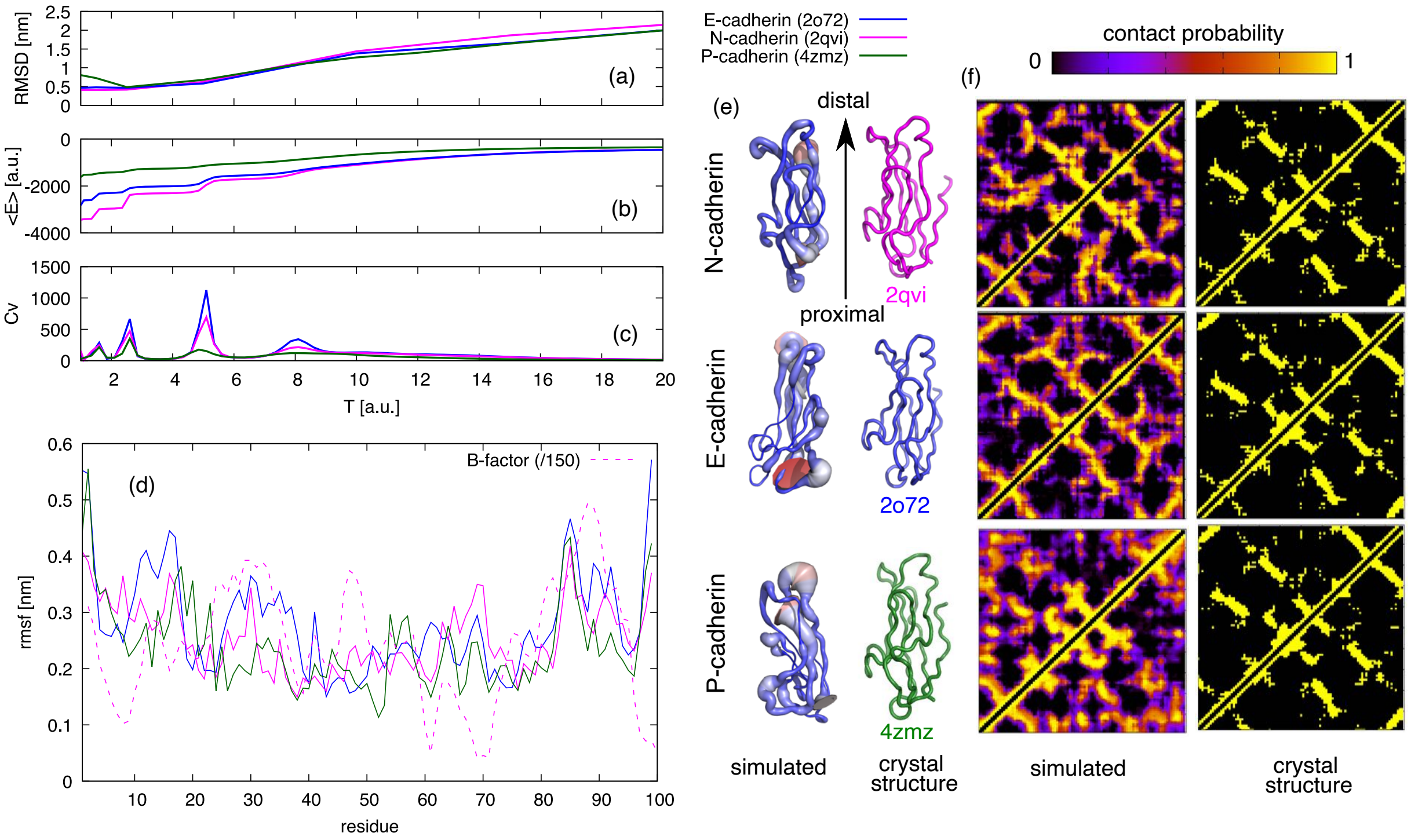

Figure 1: Simulations of monomeric EC1 domains of N-, E- and P-cadherins. The average RMSD to the crystallographic structures (a), the average energy (b) and the specific heat (c) with respect to simulation temperature (in energy units) are displayed for the three monomers. (d) The conformational fluctuations of the simulated proteins calculated at T=3.6, compared to the b-factors (dashed line) of the monomeric N-cadherin 1NCJ. (e) The equilibrium structures obtained from the simulations, in which the thickness of the cartoon reflects the fluctuations of the corresponding monomers (left side), compared to the crystallographic structures (right side). (f) the contact probabilities obtained from the simulations at T=3.6, compared to the crystallographic contact maps.

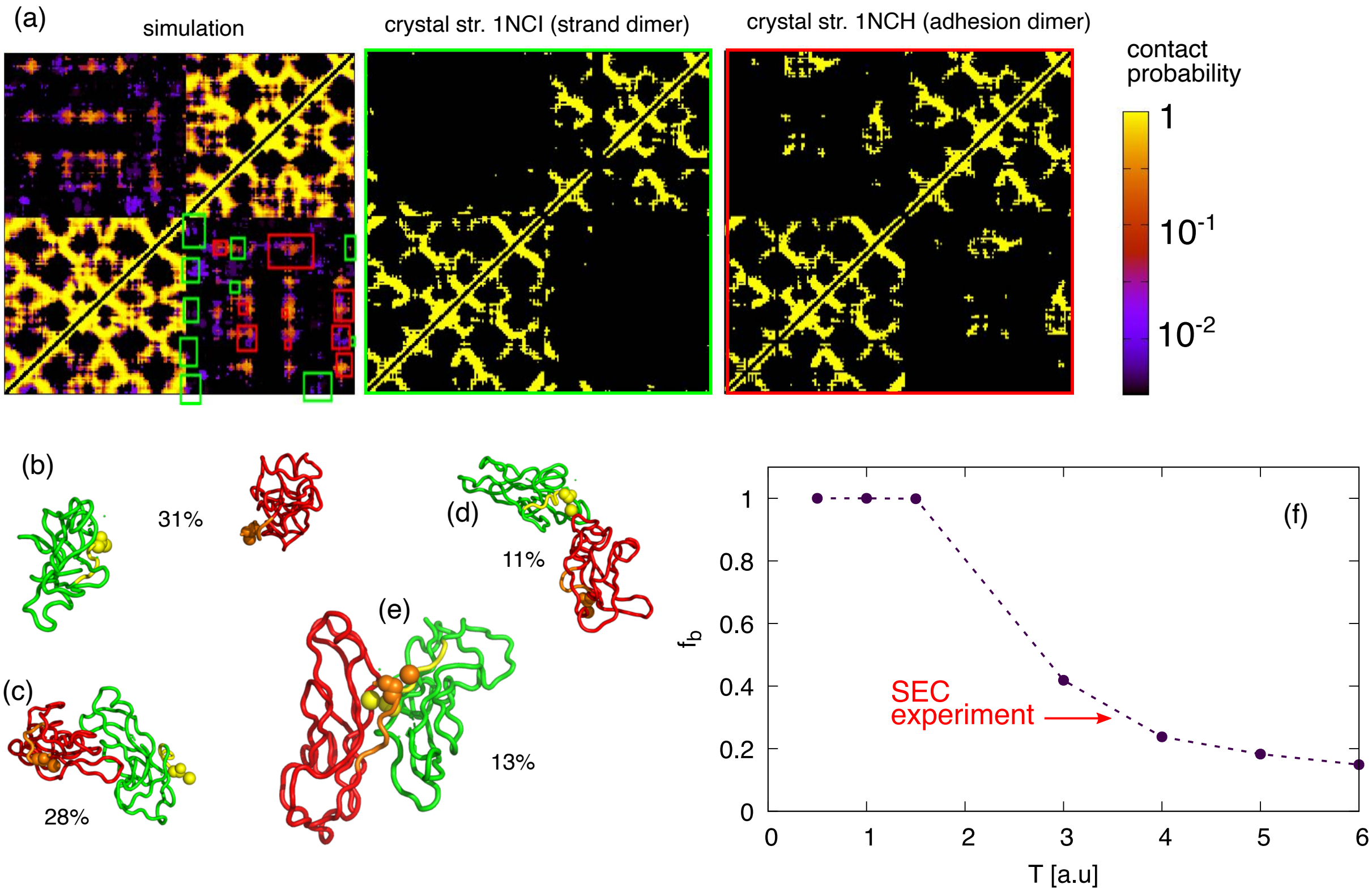

Figure 2: The result of simulations of two EC1 domains of N-cadherin at T=3.6. (a) The calculated average contact map compared with that of crystal structures 1NCI and 1NCH. The colored boxes highlight the intra-chain contacts present in the crystal structures (1NCI in green and 1NCH in red). Some representative structures obtained from a clustering of the trajectories are the two disjoint chains (b), a conformation resembling the adhesion dimer (c), a conformation bound at the ends (d) and a conformation resembling the domain-swapped strand dimer (e). The associated percentages indicate the fraction of the trajectory in each cluster. (f) The fraction $f_b$ of bound monomers as a function of temperature. The red arrow indicates the experimental value.

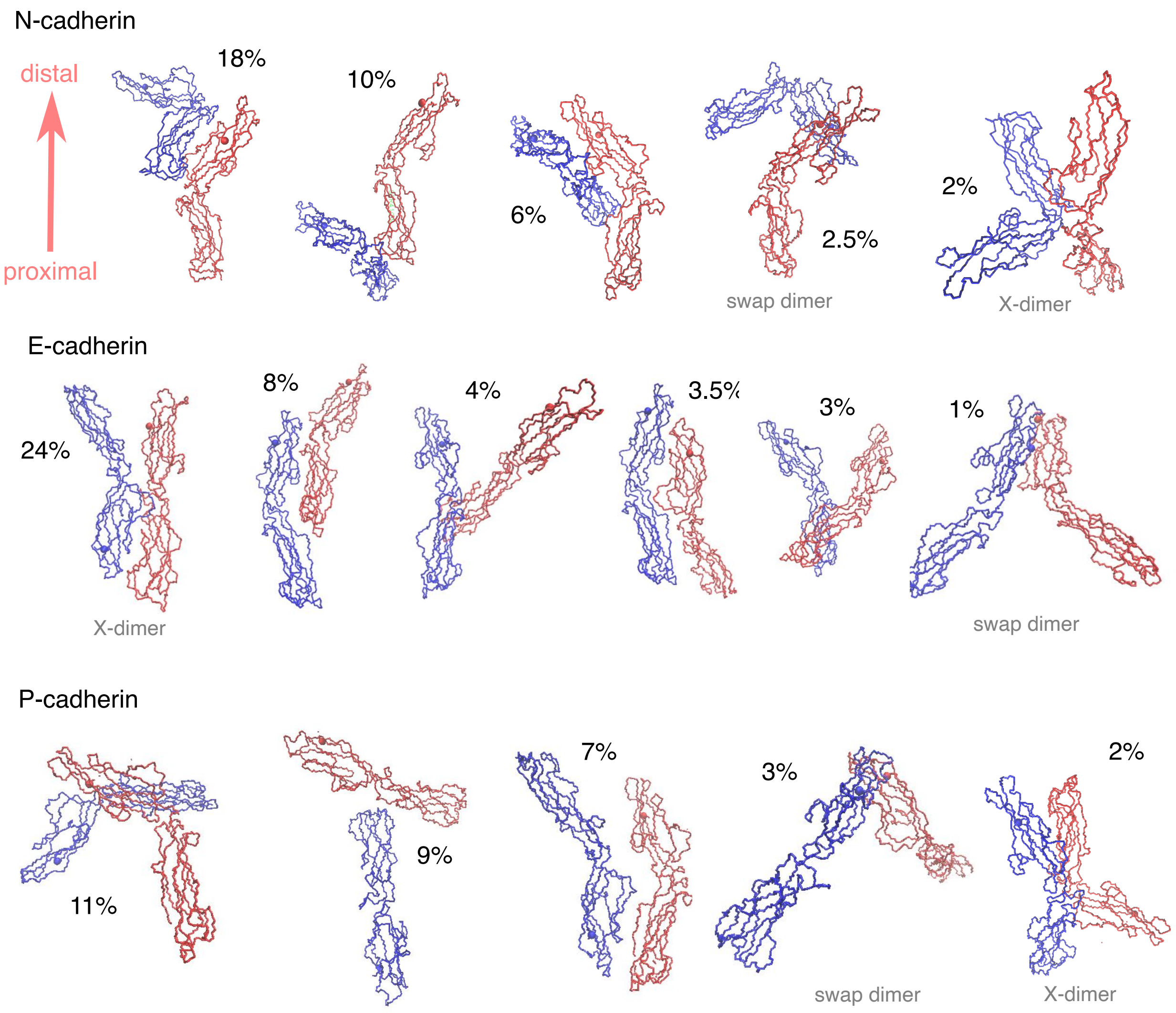

Figure 3: representative conformations of clusters obtained from the simulation of the EC1-2 dimer of N-. E- and P-cadherin. For each representative the percentage of conformations populating that cluster is indicated. The red monomers is oriented to display its distal end upward.

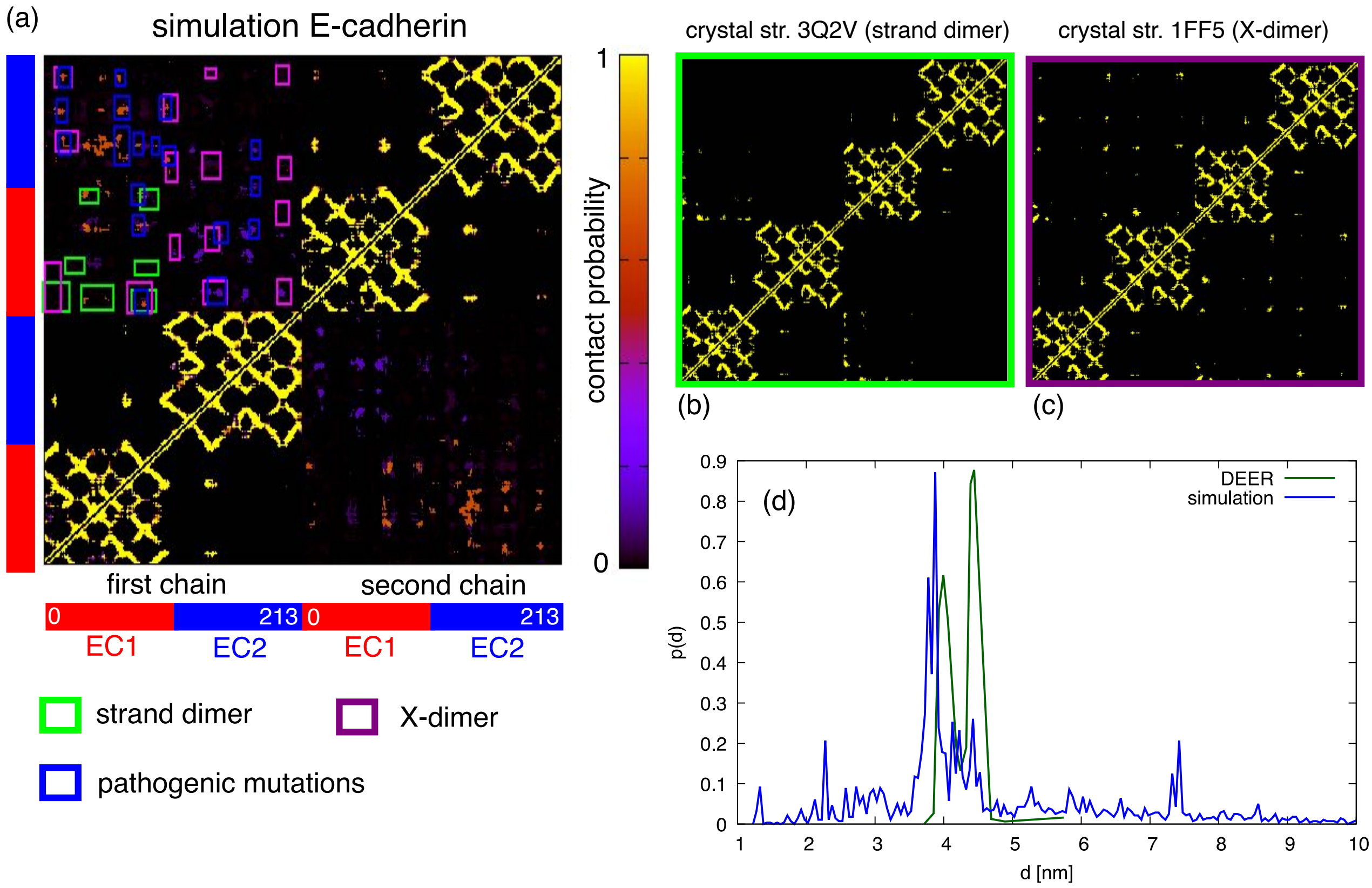

Figure 4: (a) simulated contact map of E-cadherin; the lower-left and upper-right quadrants display intra-chain contacts, the upper-left and lower-right quadrant display contacts between the two chains. (b) As a reference, we displayed the contact maps of the crystal structures of the strand dimer and of the X-dimer, and their contacts (with probability larger than 0.1) are reported with green and purple boxes, respectively, in the simulated map. The blue boxes indicate contacts between residues whose mutation is observed in tumoral tissues. (d) Comparison between the distribution of distances between residues 73-75 and residues 114-116 calculated from the simulation (in blue) and measured by DEER.

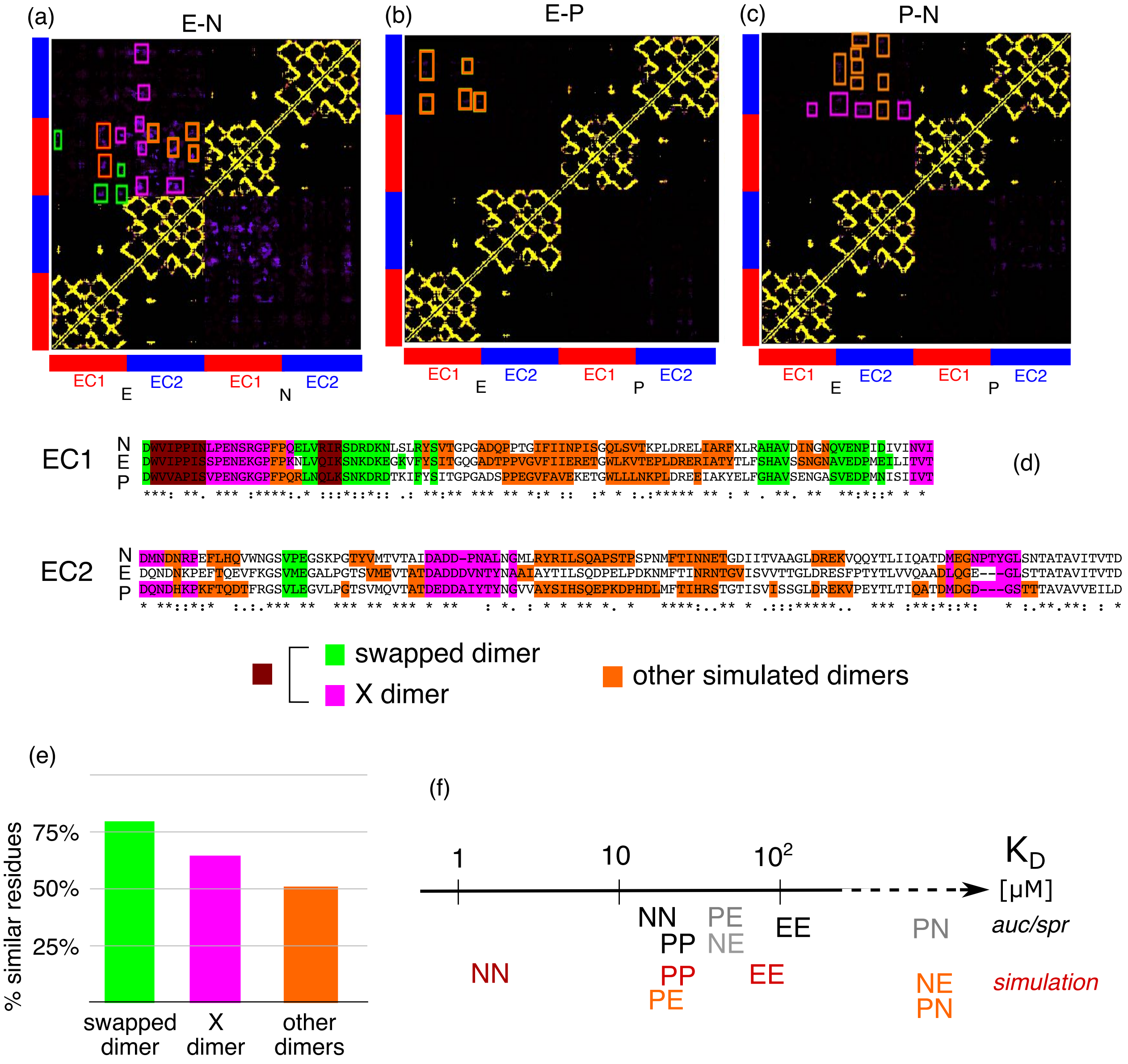

Figure 5: The average contact maps obtained simulating the EC1-2 domains of hybrid systems composed of an E- and a -N cadherin (a), of an E- and a P-cadherin (b) and a N- and P-cadherin (c). The colored boxes indicate the contacts if the swapped dimer (in green), of the X dimer (in purple) and of the other dimeric conformations sampled in the simulation (in orange). (d) A comparison of the sequences of the three cadherins, in which identical residues are marked with a star and chemically similar residues are marked with a colon. (e) The percentage of residues associated with the dimeric structures that are similar in the three proteins. (f) The experimental and the simulated dissociation constants (cf. also Table S1) for the various systems under study.

**Supplementary Materials**

Molecular recognition between cadherins studied by coevolutionary potential

Sara Terzoli and Guido Tiana

## S1 The pseudolikelihood approximation

Given an alignment $\{\sigma_i^b\}$ of $B$ homologous sequences and the empirical frequencies $f_i(\sigma)$ and $f_{ij}(\sigma,\pi)$ obtained from $\{\sigma_i^b\}$, the parameters $h_i(\sigma)$ and $J_{ij}(\sigma,\pi)$ of a Potts-like model of interaction

$$U(\{\sigma_i^b\}) = \sum_i h_i(\sigma_i) + \sum_{i<j} J_{ij}(\sigma_i,\sigma_j) \qquad \text{(S1)}$$

are obtained maximizing the log-pseudolikelihood

$$l_{ps} = \frac{1}{B}\sum_b log\left(\sum_\pi exp\left[h_i(\pi) + \sum_{j\neq i} J_{ij}(\pi,\sigma_j)\right]\right) - \sum_i \sum_\pi f_i(\pi)h_i(\pi) - \sum_{i\neq j}\sum_{\pi\tau} f_{ij}(\pi,\tau)J_{ij}(\pi,\tau) \qquad \text{(S2)}$$

as discussed in ref. 1. The maximization is constrained by two regularizers

$$\lambda \sum_{i\sigma} h_i(\sigma)^2 + \alpha \sum_{ij\sigma\pi}\left(J_{ij}(\sigma,\pi) - \varepsilon(\sigma,\tau)\right)^2, \qquad \text{(S3)}$$

where $\varepsilon(\sigma,\tau)$ is some *a priori* energies we have for the system. Calling $\delta J_{ij}(\sigma,\pi) = J_{ij}(\sigma,\pi) - \varepsilon(\sigma,\tau)$ and substituting in Eq. (S2) one can realize that this procedure is equivalent to finding the energy corrections $\delta J_{ij}(\sigma,\pi)$ to the $\varepsilon(\sigma,\tau)$ under the standard $l_2$ regularizer centered around zero.

## S2 Tuning of the energy metaparameters

To find realistic values for $\alpha$, $\varepsilon_0$ and $\varepsilon_{dih}$, we have explored their parameter space using bovine pancreatic trypsin inhibitor (BPTI, pdb code 1BPI) instead of cadherins themselves, because of the much smaller size of the former.

First, we required that the monomeric protein stays in the crystallographic native conformation at low temperature. Figure S3 displays the average RMSD with respect to the native conformation of BPTI as a function of temperature, calculated for different choices of the Lagrange multiplier $\alpha$ that controls the $l_2$ normalizer and of the parameter $\varepsilon_0$ that sets the zero of the *a priori* statistical potential (cf. Eq. 3). The Lagrange multiplier $\lambda$ on the fields (cf. Eq. S3) is kept fixed at the value 0.1, defined optimal in ref. 2, and the potential on the dihedrals is switched off. For $\alpha=10^{-5}$ and $\varepsilon_0=-1$, the average RMSD stays at the lowest value of ≈0.25 nm in a large temperature interval; thus, we used these values in the rest of the work.

Then, we performed the same kind of simulations applying the potential on the dihedrals of the backbone and varying $\varepsilon_{dih}$. The results are reported in Fig. S4. The lowest average RMSD is obtained for $\varepsilon_{dih}=90$, and consequently we used this value for the rest of the calculations. To investigate the possibility that at $\varepsilon_{dih}=90$ the properties of the chain are only controlled by the dihedral term, we performed a control simulation (gray symbols in Fig. S4) in which the parameters of the two-body interaction are randomly reshuffled, while the dihedral terms are the correct ones. As expected, the average RMSD of the protein increases drastically.

Finally, we performed a parallel-tempering simulation on BPTI and on another protein, acyl-coenzime A binding protein (ACBP, pdb code 2ABD) on a wider range of temperature to study the folding transition of test proteins within the model. In this range of temperatures, both proteins display a marked increase in the average RMSD (cf. Fig. S5a). The specific heat of BPTI, calculated from the multiple-histogram method[3], displays a low-temperature peak around T=5, corresponding to freezing of the chain in the lowest-energy state (cf. Fig. S5b), a denaturation peak around T=10 and a small peak associated to swelling at larger temperature. ACBP displays the same freezing peak and a much broader peak associated with denaturation and swelling.
Again, to study the role of the different terms in the potential, we plotted in Fig. S5c the average two-body energy and the average dihedral energy. Upon increasing the temperature, the two-body term follows closely the total energy, while the dihedral term displays milder changes. Moreover, the two terms seem poorly correlated in the visited conformations (cf. Fig. S5d). These data suggest that the dihedral term gives a non-trivial, but important (cf. Fig. S4), contribution to the stabilization of the proteins.

**S3 References**

1. Ekeberg M, Lövkvist C, Lan Y, Weigt M, Aurell E. Improved contact prediction in proteins: Using pseudolikelihoods to infer Potts models. Phys. Rev. E, Stat. nonlinear, soft matter Phys. 2013;87:620630.
2. Fantini M, Malinverni D, De Los Rios P, Pastore A. New Techniques for Ancient Proteins: Direct Coupling Analysis Applied on Proteins Involved in Iron Sulfur Cluster Biogenesis. Front. Mol. Biosci. 2017;4:40.
3. Ferrenberg A, Swendsen R. Optimized Monte Carlo data analysis. Phys. Rev. Lett. 1989;63:1195–1198.
4. Vendome J, Felsovalyi K, Song H, Yang Z, Jin X, Brasch J, Harrison OJ, Ahlsén G, Bahna F, Kaczynska A, et al. Structural and energetic determinants of adhesive binding specificity in type I cadherins. Proc. Natl. Acad. Sci. U. S. A. 2014;111:E4175–E4184.

| System | $K_D$ simulation | $K_D$ experiment[4] |
|---|---|---|
| NN | 2.0 | 26 |
| PP | 35.1 | 31 |
| EE | 81.5 | 96 |
| PE | 14.0 | ~50 |
| NE | >100 | ~50 |
| PN | >100 | >100 |

Table S1: comparison between the dissociation constants (in $\mu$M) obtained from the simulation and those obtained by analytical ultracentrifugation or plasmon resonance experiments for the EC1-2 system of different cadherins. The $K_D$ is calculated from the simulations counting the fraction of frames in which the two chains contacts (considering a lower threshold on inter-chain contact energies of -26 to define bound conformations) and knowing that the side of the box is 20 nm.

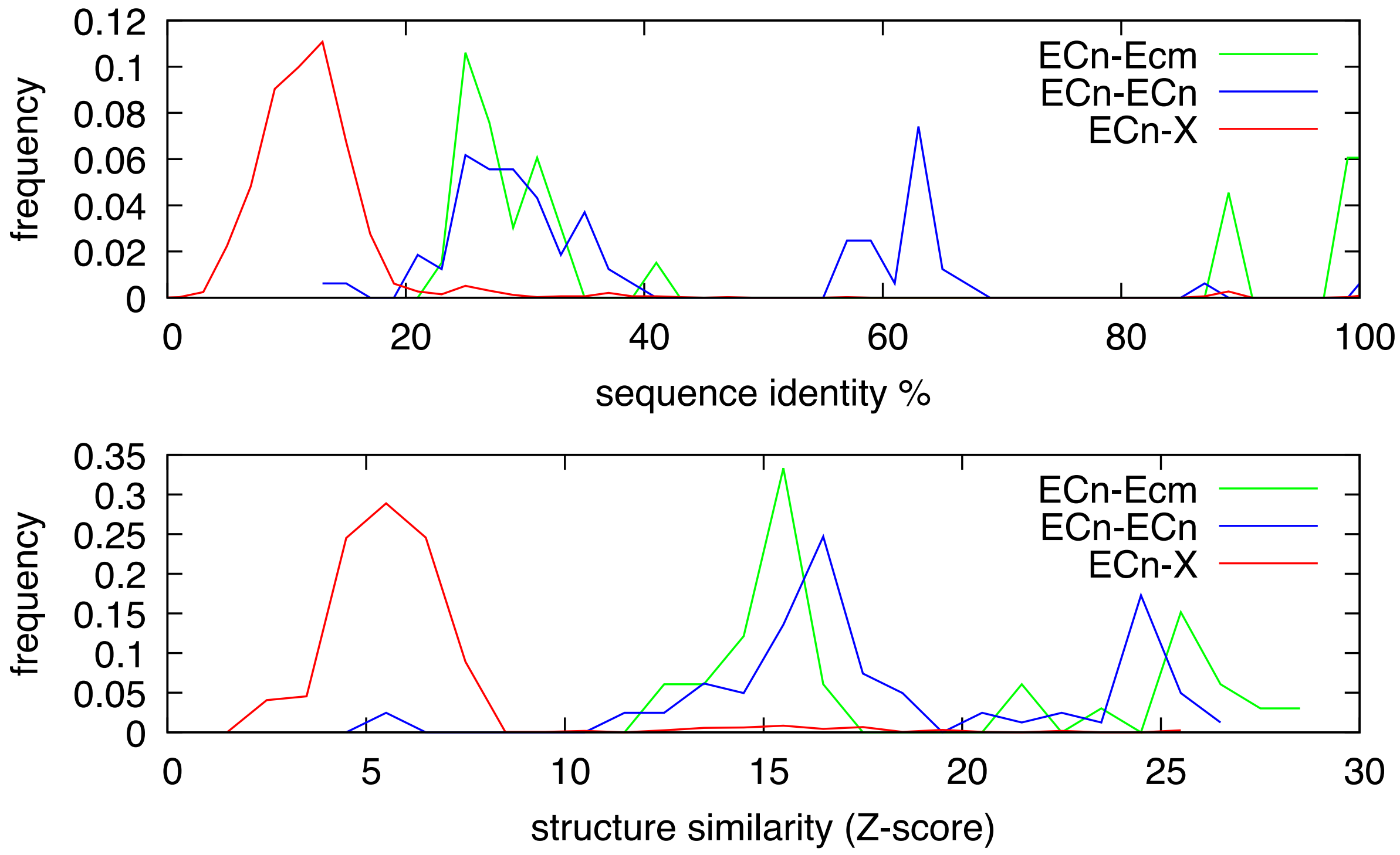

Figure S1: Sequence (above panel) and structural (below panel) similarity between the same domain of different cadherins, between different domains of the same cadherin type and between cadherin domains and, as a control (X), between cadherin domains and structurally similar domains (Z>2), as calculated from the Dali database ( http://ekhidna2.biocenter.helsinki.fi/dali/ ).

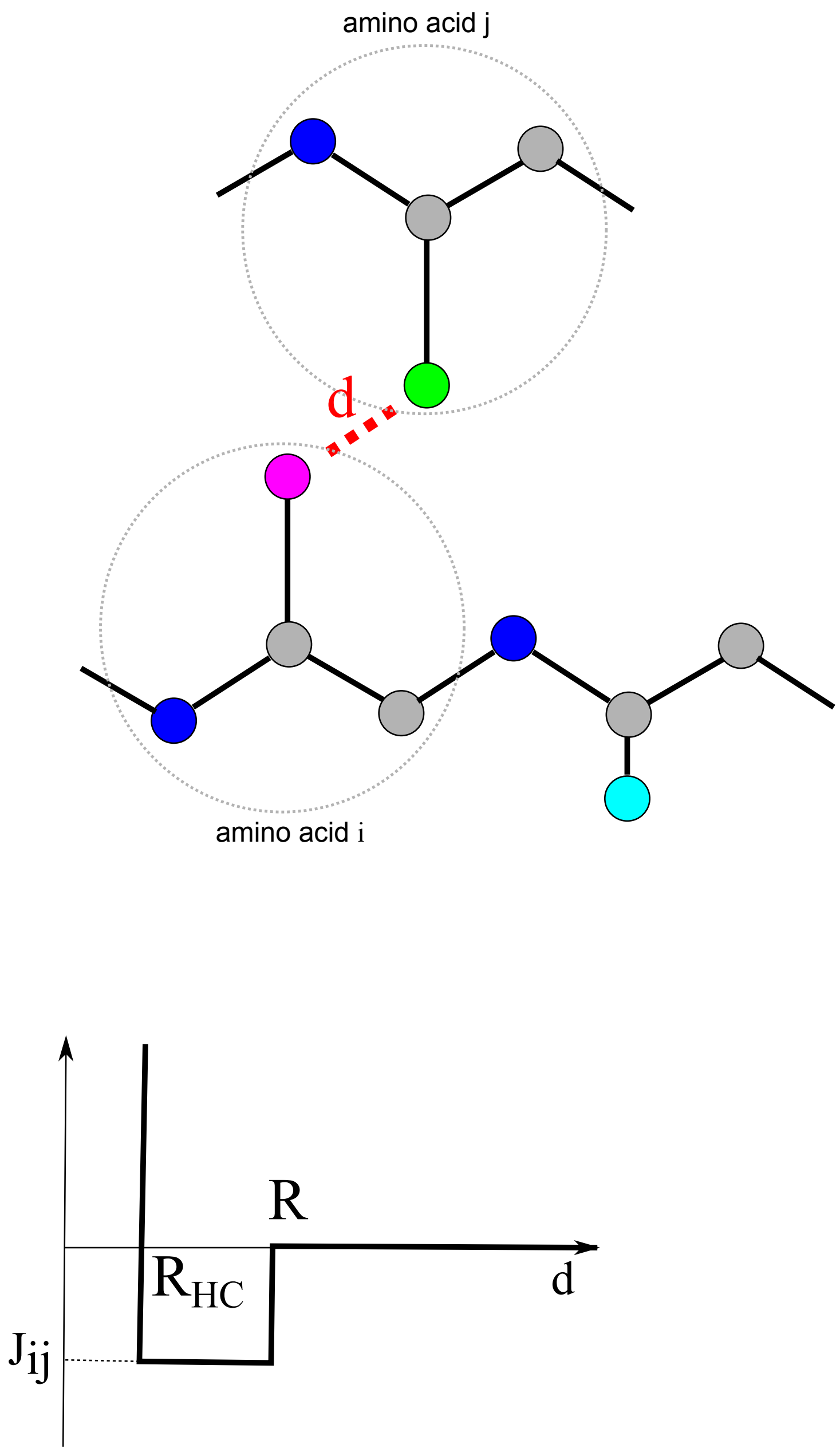

Figure S2: a sketch of the model. Each amino acid is described by four beads, representing the atoms N, CA, C and the side chain, respectively. The interaction between two amino acids depends on the distance *d* between the side chains, and has the form of a spherical well with range R, hard-core radius $R_{HC}$ and depth $J_{ij}$ depending on the specific pair. Each pair of atoms have a hard-core repulsion.

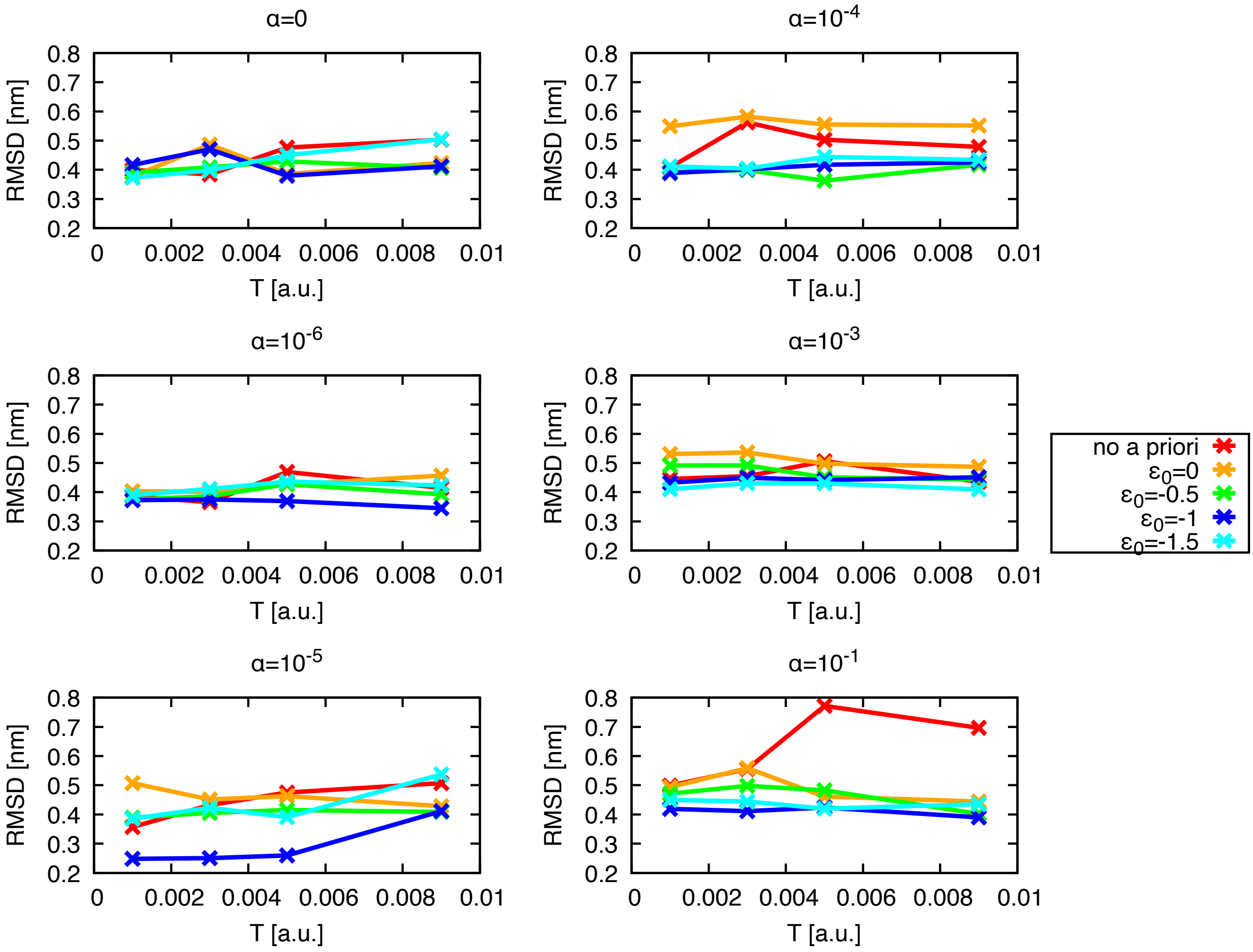

Figure S3: the average RMSD of BPTI as a function of temperature T, obtained at various values of the Lagrange multiplier $\alpha$ that controls the $l_2$ normalizer and of the parameter $\varepsilon_0$ that sets the zero of the *a priori* statistical potential. In the case with "no a priori" we used a $l_2$ normalizer centered at zero.

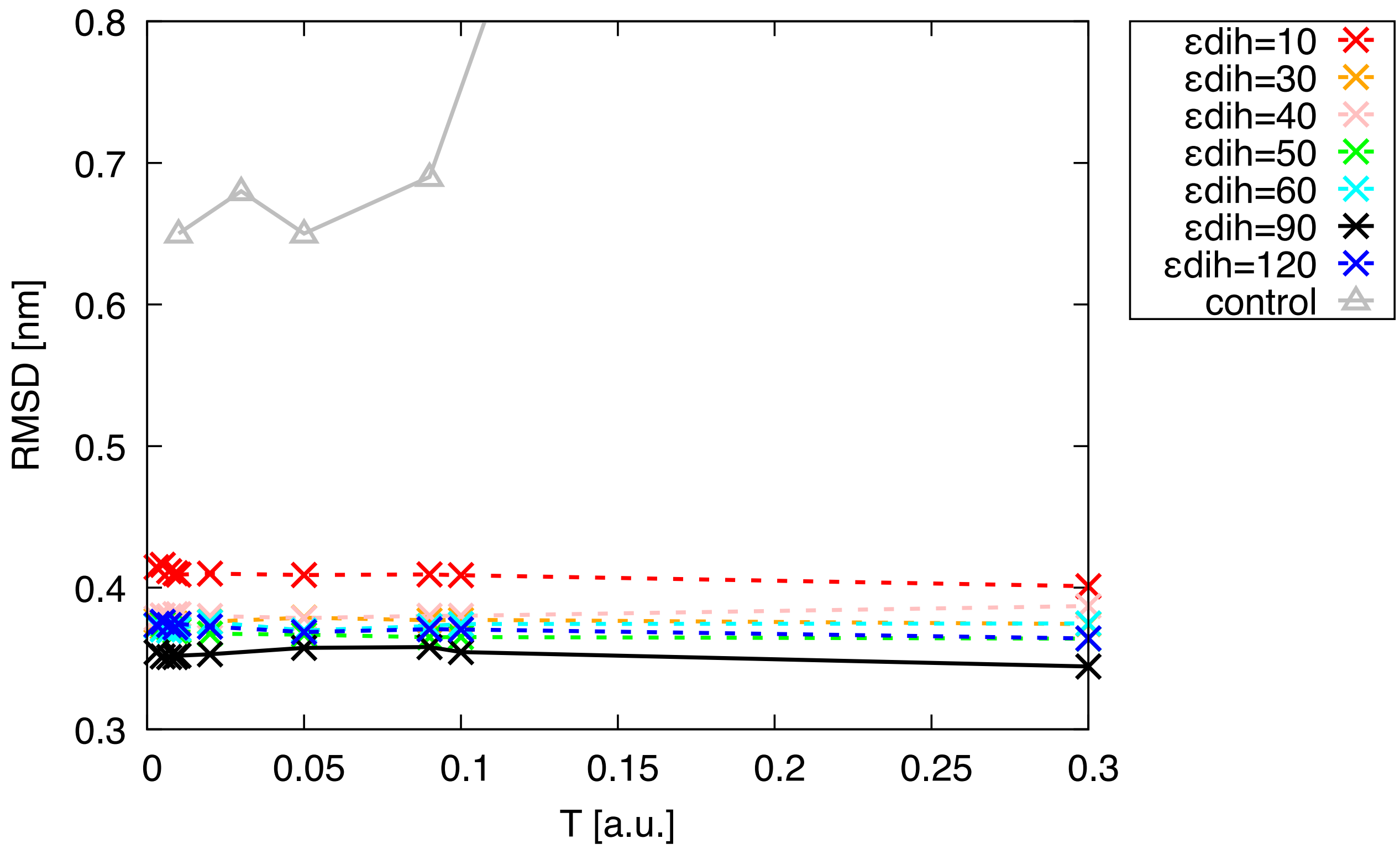

Figure S4: the average RMSD of BPTI as a function of temperature T using $\alpha=10^{-5}$ and $\varepsilon_0=-1$, for different choices of $\varepsilon_{dih}$, that controls the potential on the dihedrals. The control points are obtained reshuffling at random the parameters of the two-body potentials, and keeping the correct dihedral potential with $\varepsilon_{dih}=90$.

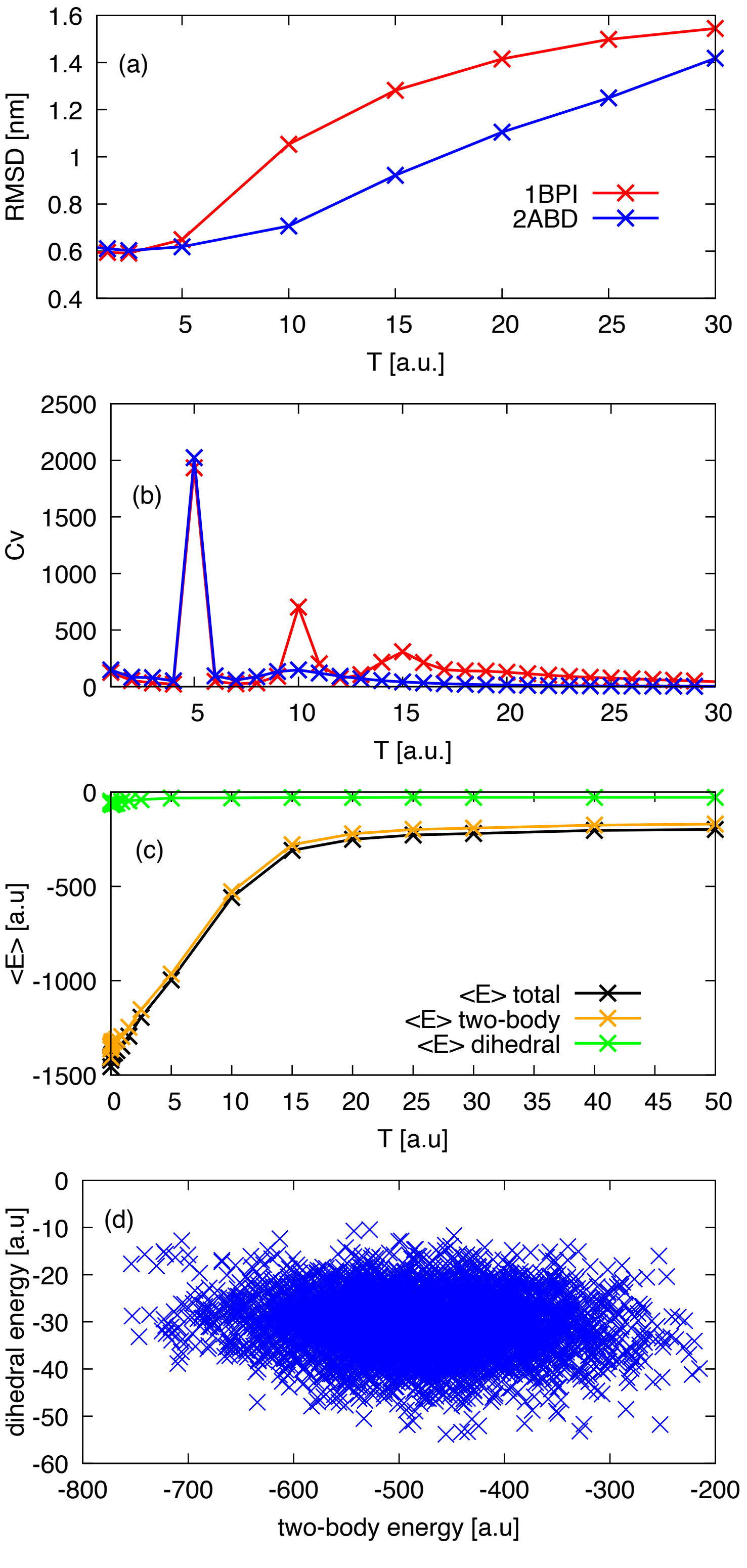

Figure S5: The average RMSD (a) and the heat capacity $C_V$ (b) of BPTI and ACBP as a function of temperature, calculated from parallel-tempering simulations. (c) The two-body and the dihedral terms of the total energy for BPTI at various temperatures. (d) A scatter plot of the two-body and of the dihedral terms of the energy for conformations at T=5.

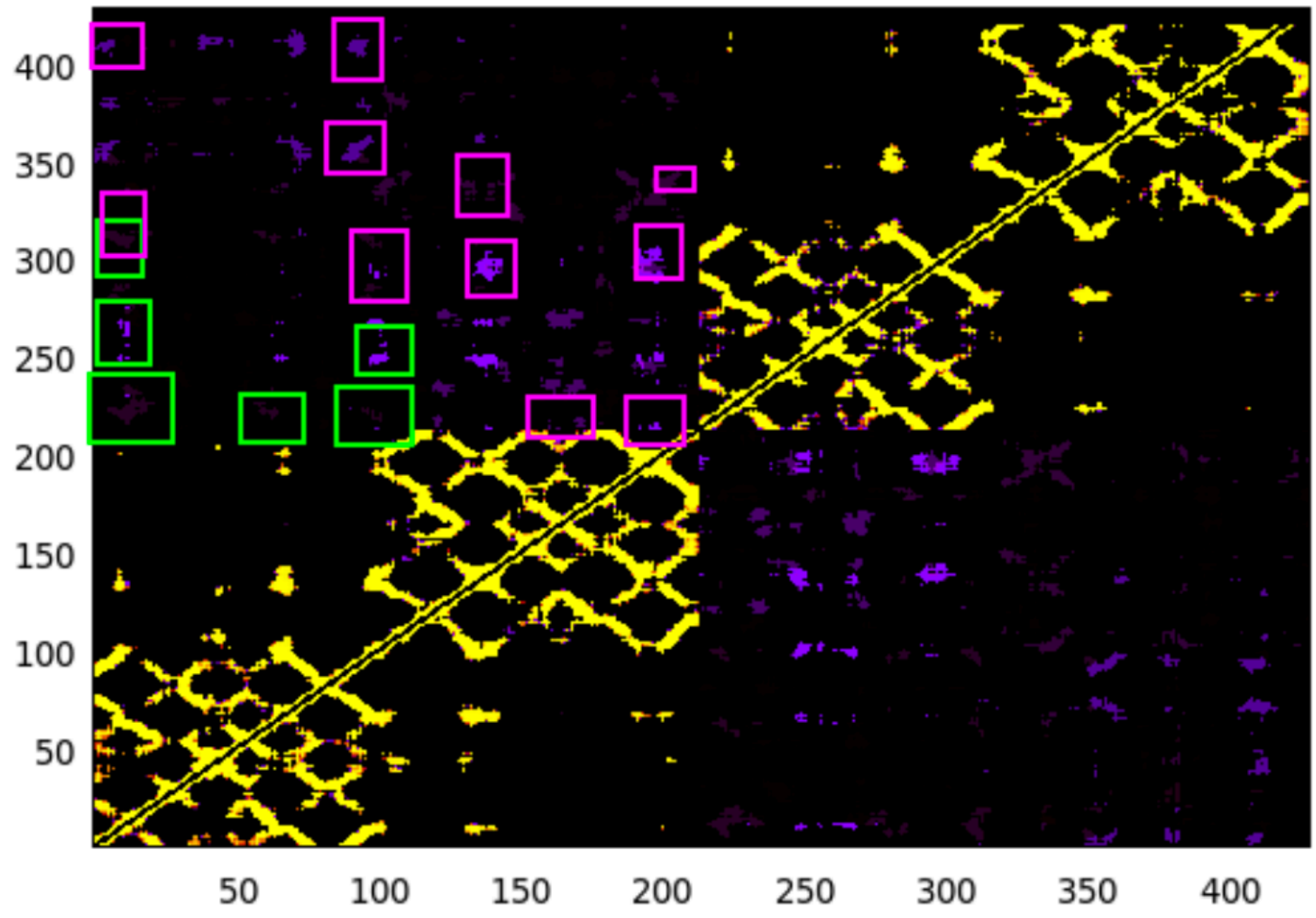

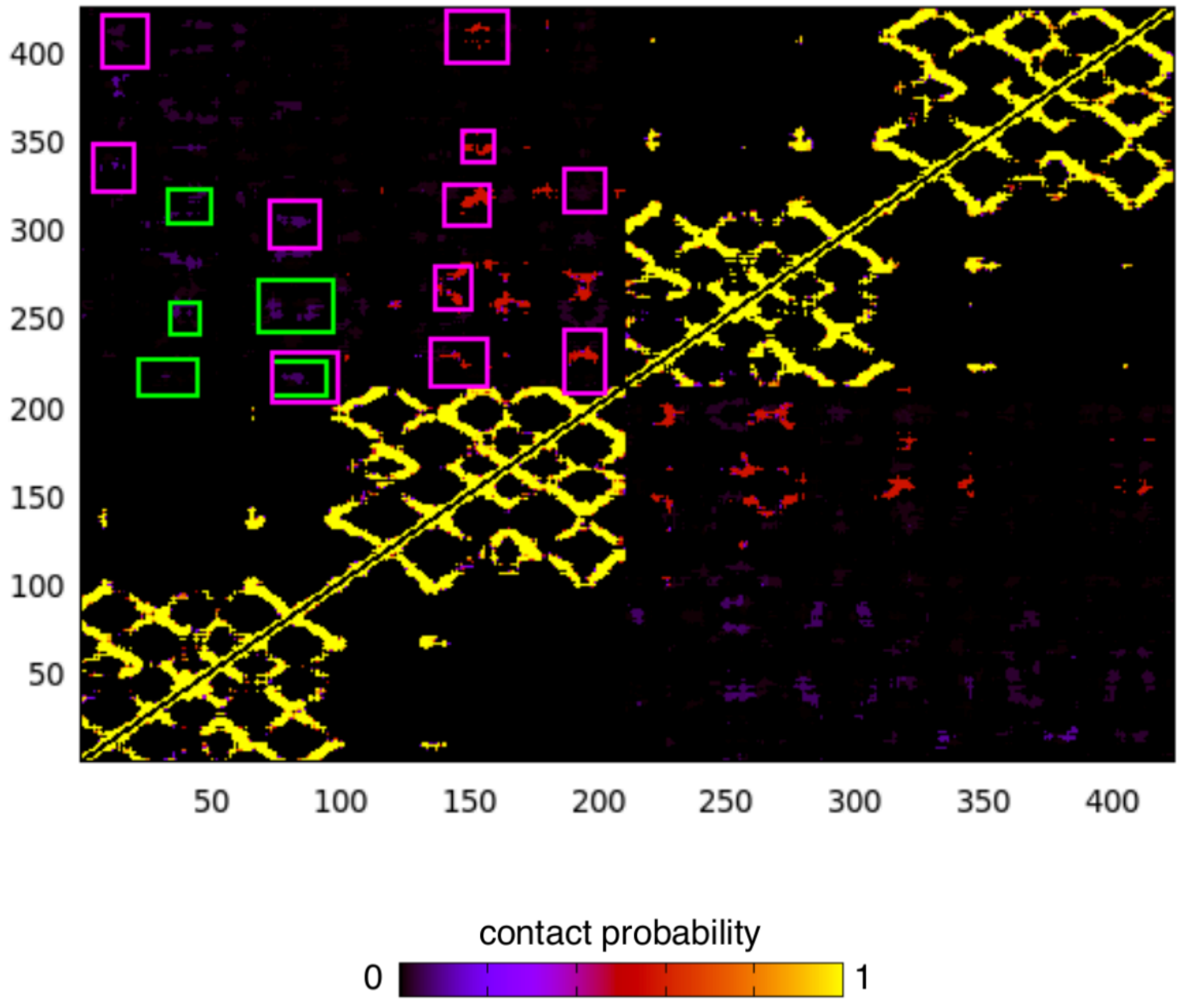

Figure S6: the simulated contact maps of N-cadherin (above) and P-cadherin (below). Green boxes mark contacts present in the strand dimer, purple boxes those present in X-dimers.

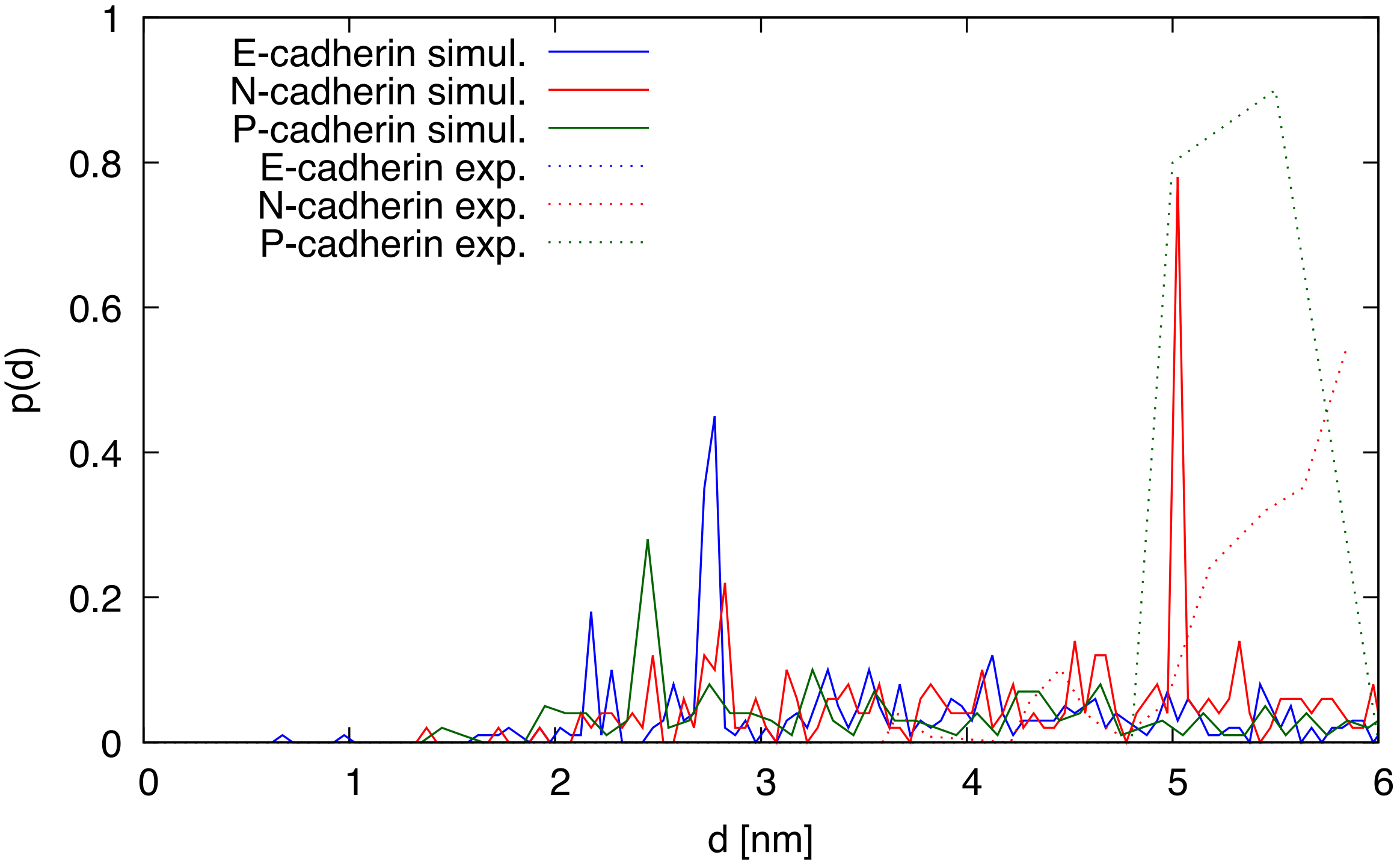

Figure S7: comparison of the simulated distribution of distances between residues 135 with the results of DEER experiments.